\pdfoutput=1
\documentclass{ifacconf}

\usepackage{graphicx}      
\usepackage{natbib}        
\usepackage{xcolor}
\usepackage{mathtools}
\usepackage{makecell}
\usepackage{amssymb}
\usepackage[utf8]{inputenc}
\usepackage{ACSD-CR}

\usepackage{xspace}
\usepackage{soul}
\usepackage{mathtools}
\usepackage{makecell}
\usepackage{nicefrac}
\usepackage{float}
\usepackage{wrapfig}

\newcommand{\ie}{\unskip, i.\,e.,\xspace}
\newcommand{\eg}{\unskip, e.\,g.,\xspace}

\definecolor{dgreen}{rgb}{0.0, 0.5, 0.0}

\usepackage{textcomp}

\makeatletter
\newcommand{\subalign}[1]{%
	\vcenter{%
		\Let@ \restore@math@cr \default@tag
		\baselineskip\fontdimen10 \scriptfont\tw@
		\advance\baselineskip\fontdimen12 \scriptfont\tw@
		\lineskip\thr@@\fontdimen8 \scriptfont\thr@@
		\lineskiplimit\lineskip
		\ialign{\hfil$\m@th\scriptstyle##$&$\m@th\scriptstyle{}##$\crcr
			#1\crcr
		}%
	}
}

\begin{document}

\begin{frontmatter}

\title{A method of online traction parameter identification and mapping} 

\author[First]{Alexander Kobelski} 
\author[First]{Pavel Osinenko} 
\author[First]{Stefan Streif}

\address[First]{Technische Universität Chemnitz, Automatic Control and System Dynamics Laboratory, 
   Germany (e-mail: \{alexander.kobelski;pavel.osinenko;stefan.streif\}@etit.tu-chemnitz.de)}

\SETCR{\CRIFAC{https://doi.org/10.1016/j.ifacol.2020.12.909}{IFAC-PapersOnLine}{Alexander Kobelski, Pavel Osinenko, Stefan Streif}}
\begin{abstract}                
Fuel consumption of heavy-duty vehicles such as tractors, bulldozers etc. is comparably high due to their scope of operation.
The operation settings are usually fixed and not tuned to the environmental factors, such as ground conditions.
Yet exactly the ground-to-propelling-unit properties are decisive in energy efficiency.
Optimizing the latter would require a means of identifying those properties.
This is the central matter of the current study.
More specifically, the goal is to estimate the ground conditions from the available measurements, such as drive train signals, and to establish a map of those.
The ground condition parameters are estimated using an adaptive unscented Kalman filter.
A case study is provided with the actual and estimated ground condition maps.
Such a mapping can be seen as a crucial milestone in optimal operation control of heavy-duty vehicles.
\end{abstract}

\begin{keyword}
Traction control, Identification algorithms, Data storage, Kalman filters, Dynamic modelling, Vehicle Dynamics
\end{keyword}

\end{frontmatter}


\section{Introduction}
Increasing fuel costs due to progressing depletion of fossil resources put ever stronger requirements on the productivity, energy efficiency and related characteristics of heavy-duty vehicles.
Optimization of the vehicle operation may be realized by various factors, such as implement positioning, tire pressure adjustment, traction and engine control.
	But one thing is common: determination of an optimal set point has to account for the properties of the ground-to-propelling-unit properties.

\begin{figure}
	\centering
	\includegraphics[width=0.9\linewidth]{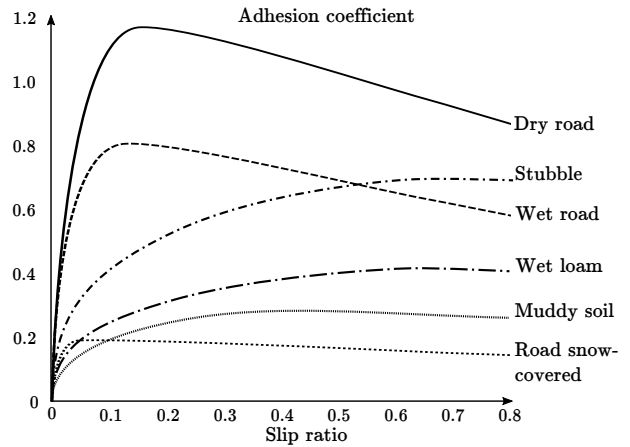}
	\caption{Typical $\mu(s)$-characteristics on different soils.}
	\label{fig:EE_AC_slip_graph}
\end{figure}

Primarily, the ground conditions influence the traction dynamics.
These can be characterized by two factors -- the energy efficiency and the adhesion coefficient, which equals the propulsion force normalized by the vehicle's weight.
Both can be considered as functions of the wheel slip ratio (see Fig.~\ref{fig:EE_AC_slip_graph} for some adhesion-slip-curve examples on different soils).
A detailed description of traction-slip characteristics may be found in \cite{Sohne1964}.
High operation performance requires balancing the propulsion and the energy efficiency.
This is complicated precisely due to the lack of online knowledge of the adhesion-slip curve.

So far, the methods of operation optimization, which account for the adhesion-slip relation, are mostly offline. 
The techniques of the so-called traction prediction \citep{Schreiber2008,Battiato2017} rely on certain empirical parameters of the propelling unit, chassis, type of soil etc. 
The user can thereby calculate an operation set point for a particular situation and adjust the settings accordingly, in a way similar to a look-up table.
An immediate disadvantage of these methods is their offline nature \ie they do not account for changing ground conditions.
	On the other hand, the online methods do not use identification of the adhesion-slip characteristic.
In particular, \citet{Reichensdoerfer2018} proposed a nonlinear control design based on input-output linearizion that takes oscillatory behavior of the powertrain into account.
\citet{Ishikawa2012} suggested usage of GPS antennae to measure speed, calculate the slip ratio and adjust the implement position when the slip ratio exceeds a given threshold.
It should be noted that the problems of traction control are also relevant in railway vehicles \citep{Novak2018}.
Neither of the above use identification of ground properties to adapt the control parameters.



However, some empirical indices, such as the one based on the Brixius model \citep{Brixius1987}, started to find applications.
For instance, \citet{Kim2018} used it in a slip controller.
\cite{Addison2018} proposed using a data buffer coupled with an parameter optimization algorithm for traction parameter identification.
\cite{Pentos2017} used an artificial neural network to predict the influence of the soil texture, soil moisture and compaction etc. on the propulsion force and traction efficiency.
\cite{Rajamani2012-adhesion-est} developed observers to estimate friction coefficients of individual wheels during operation from various measurements \eg engine torque, brake torque and GPS measurements.
Similarly \cite{Wang_FrictionEstimation} developed a real-time tire-road friction coefficient measurement system which uses a differential GPS and a nonlinear longitudinal tire force model.
\cite{Turnip2013-adhesion-est-EKF} used an identification approach based on the extended Kalman filter (EKF),
whereas \cite{Hamann2014-adhesion-est-UKF} suggested to use a superior variant of the EKF -- the unscented Kalman filter (UKF).
The UKF \citep{Wan2000-UKF,VanDerMerwe2004-UKF} will also be used in this work.

\begin{figure}[h]
	\centering
	\includegraphics[width=0.8\linewidth]{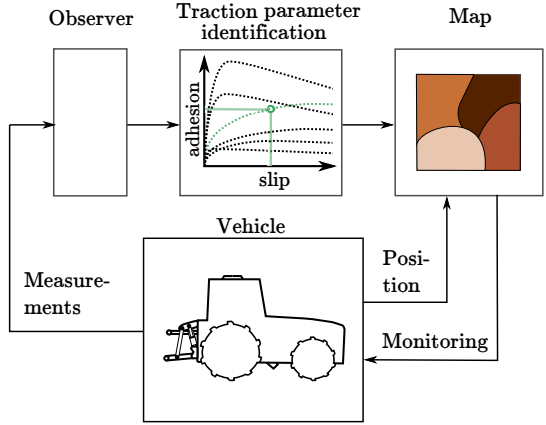}
	\caption{General flowchart of the ground condition identification and mapping. 
}
	\label{fig:program_flow_simple}
\end{figure}

It can be observed from the literature, that the problem of ground condition identification is an actively researched topic and there are many open questions.
The goal of this work is to develop means of online identification and mapping of those condition parameters.
It continues the methodology developed in \cite{Osinenko2014,Osinenko2016,Osinenko2017}.
An adaptive UKF is used for the identification of the ground-to-propelling-unit properties, namely, the adhesion and the rolling resistance coefficients. 
The slip-adhesion characteristic, or simply $\mu(s)$-characteristic, is estimated from a single operating point.
The way this is done is to use the model suggested in \cite{Osinenko2017}, 
and to reduce the number of its free parameters to a single one (see details in Sections \ref{sec:traction_par_identification} and \ref{sec:mapping}).
The traction parameters, as well as the said parameter characterizing the slip-adhesion relation, are mapped into a look-up table using GPS data recorded during operation.
Data are extrapolated/interpolated using a distance\textendash based weighing algorithm.
Fig. \ref{fig:program_flow_simple} shows a general framework.
To elaborate it, first the modeling matters are discussed in the next Section \ref{sec:traction_dynamics}.


\section{Traction dynamics}
\label{sec:traction_dynamics}

\begin{figure}
	\centering
	\includegraphics[width=0.7\linewidth]{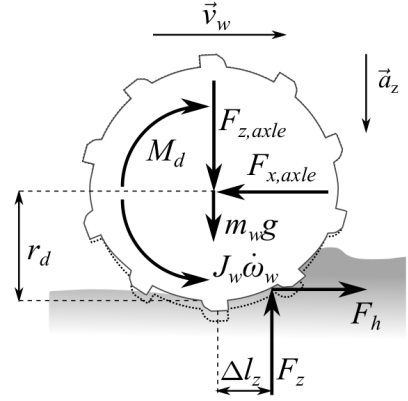}
	\caption{Diagram depicts forces and torque at the wheel. 
	}
	\label{fig:wheel_forces}
\end{figure}

Consider the wheel force diagram of Fig.~\ref{fig:wheel_forces}.
There, the vertical, or normal, force $F_\mathrm{z}$ is the sum of axle load $F_\mathrm{z,axle}$ and wheel weight plus any vertical acceleration, see \eqref{eq:force_balance_vertical}.
The motor causes a driving torque $M_\mathrm{d}$ which exerts a horizontal force $F_\mathrm{h}$ at the wheel.
The dynamical rolling radius  $r_\mathrm{d}$ is defined to describe the deformed wheel's distance between its center and bottom.
Due to the tire and soil deformation the point of application of the soil reaction forces is shifted by $\Delta l_\mathrm{z}$.
By convention the term $\Delta l_\mathrm{z}F_\mathrm{z}$ is assumed to be equal $r_\mathrm{d}F_\mathrm{t}$, where $F_\mathrm{t}=\rho_\mathrm{t}F_\mathrm{z}$, i.e. the tire-deformation rolling resistance.
The revolution speed of the wheel $\dot{\omega}_\mathrm{w}$ can now be calculated using a torque balance approach, see \eqref{eq:torque_balance}.
The reaction of the vehicle body $F_\mathrm{x,axle}$ is exerted in the opposite direction of $F_\mathrm{h}$.
The rolling resistance is divided in the internal rolling resistance $F_\mathrm{t}$ with its coefficient $\rho_\mathrm{t}$ caused by the deformation of the tire and the external rolling resistance $F_\mathrm{s}$ with its coefficient $\rho_\mathrm{s}$ caused by the deformation of the soil.
The force and and torque balance yields:
\begin{align} 
m_\mathrm{w}\dot{v}_\mathrm{w} 			&= F_\mathrm{h} - F_\mathrm{x,axle} - \rho_{\mathrm{s}}F_\mathrm{z}, \label{eq:force_balance_horizontal} \\
J_\mathrm{w}\dot{\omega}_\mathrm{w} 	&= M_\mathrm{d} - r_\mathrm{d}F_\mathrm{h} - r_\mathrm{d}\rho_\mathrm{t}F_\mathrm{z}, \label{eq:torque_balance} \\
m_\mathrm{w}a_\mathrm{z} 				&= F_\mathrm{z} - m_\mathrm{w}g - F_\mathrm{z,axle}, \label{eq:force_balance_vertical} 
\end{align}
where $m_\mathrm{w}$ is the wheel mass, $v_\mathrm{w}$ is the ground speed, $J_\mathrm{w}$ is the wheels inertia, $a_\mathrm{z}$ is the vertical acceleration and $g$ is the gravitation.
The total rolling resistance is dominated by the soil-deformation resistance $F_\mathrm{s}$ and significantly influences the energy efficiency.

The adhesion coefficient $\mu$ characterizes the relation between the horizontal and vertical force as follows:
\begin{equation}\label{eq:mu_physical}
F_\mathrm{h}=\mu F_\mathrm{z}.
\end{equation}
The part of the horizontal force that actually drives the wheel forward is called net traction ratio $\kappa$
\begin{equation}\label{eq:mu_rho_kappa}
\kappa=\mu-\rho_\mathrm{s}.
\end{equation}
The following definition of slip is used:
\begin{equation}
\begin{array}{cccc}
s= & 1-\frac{|v|}{r_{d}|\omega_{w}|}, & \text{if} & |v|\leq r_{d}|\omega_{w}|,\\
s= & -1+\frac{r_{d}|\omega_{w}|}{|v|}, & \text{if} & |v|>r_{d}|\omega_{w}|.
\end{array}\label{eq:slip}
\end{equation}
It ranges from -1 (locked wheel) to 1 (spinning on the spot). 
The energy efficiency $\eta$ is defined by the formula:
\begin{equation}
\eta=\frac{\kappa}{\kappa+\rho}(1-s).\label{eq:traction-eff}
\end{equation}
Note that $\kappa$, $\mu$ and $\eta$ are functions of slip $s$.

The soil deformation rolling resistance is summarized into $F_\mathrm{s}=\sum_{i=1}^{4}F_{\mathrm{s}i}=\rho_\mathrm{s}mg$.
The vehicle dynamics in driving direction now are
\begin{equation}
m\dot{v} = \sum_{i=1}^{4}F_{\mathrm{h}i} - F_\mathrm{dx} - \rho_\mathrm{s}mg ,\label{eq:vehicle_dynamics}
\end{equation}
where $m$ is the vehicle mass, $F_\mathrm{dx}$ is the horizontal part of the drawbar pull, i.e. the implement resistance and $i$ is the wheel index.

The tire deformation resistance $\rho_\mathrm{t}$ mainly depends on tire type and inflation pressure and can therefore be estimated prior to operation.
A value of $\rho_\mathrm{t}=0.015$, as suggested by \cite{Schreiber2007}, is used for simulation studies in this work.
The soil deformation resistance $\rho_\mathrm{s}$ and $\mu$ will be estimated using a state observer, i.e. Kalman filter, see Section~\ref{sec:traction_par_identification}.
The dynamical rolling radius
\begin{equation}\label{eq:roll_radii}
r_\mathrm{d}=r_0-\Delta r,
\end{equation}
where $r_0$ is the unloaded wheel radius and  $\Delta r$ is the tire deformation, can be approximated from vertical forces using the empirical formula suggested by \cite[p.40]{Guskov1988}:
 \begin{equation}\label{eq:roll_radii_deformation}
 \Delta r=\frac{F_\mathrm{z}}{2\pi\cdot 10^5\cdot p_\mathrm{t}\sqrt{\frac{b_\mathrm{t}}{2}r_0}},
 \end{equation}
with $p_\mathrm{t}$ the tire air pressure and $b_\mathrm{t}$ the tire section width.

The parameters  $J_\mathrm{w}$, $m_\mathrm{w}$, $m$ and $r_0$ are assumed known.
For details on estimation of $F_\mathrm{z}$ and $M_\mathrm{d}$, please refer to \cite{Osinenko2015a}.
The state model is required by the state observer/parameter estimator introduced in the following Section \ref{sec:traction_par_identification}.

\section{Ground condition identification } \label{sec:traction_par_identification}

This section describes first the method of identifying the adhesion and rolling resistance coefficients, and then proceeds to the $\mu(s)$-characteristic.

\subsection{Online traction parameter identification}


The identification algorithm used in this work bases on the adaptive UKF suggested in \cite{Jiang2007}.
The details thereof are given for the sake of completeness whereas the core of the method lies with the identification of the $\mu(s)$-characteristic and mapping of the ground condition parameters.
As mentioned above, the two parameters that impact the performance of the vehicle the most are adhesion coefficient $\mu$ and soil-deformation rolling resistance $\rho_\mathrm{s}$.
For their identification, an adaptive UKF with a fuzzy-logic supervisor (AUKF-FS) is used.
Its purpose is to estimate the state $\mathbf{x}_{k}$ from the measured output $\mathbf{y}_{k}$. 
The generic model description behind the UKF reads: 
\begin{equation}
	\begin{aligned} & \mathbf{x}_{k}=f\left(\mathbf{x}_{k-1},\mathbf{u}_{k-1}\right)+\mathbf{q}_{k-1},\\
	& \mathbf{y}_{k}=h\left(\mathbf{x}_{k}\right)+\mathbf{r}_{k}.
	\end{aligned}
	\label{eq:discr-nonlin-model}
\end{equation}
Here, $\mathbf{x}_{k}\in\mathbb{R}^{n}$ is the state vector, $\mathbf{u}_{k}\in\mathbb{R}^{p}$ is the input vector, $\mathbf{y}_{k}\in\mathbb{R}^{m}$ is the output vector, $f\left(\mathbf{x}_{k-1},\mathbf{u}_{k-1}\right)$ is the non\textendash linear state model, $h(\mathbf{x}_{k})$ is the measurement model, $q_{k}\sim\mathcal{N}\left(0,\mathbf{Q}\right),r_{k}\sim\mathcal{N}\left(0,\mathbf{R}\right)$ are the state and measurement random noises with zero mean and covariance $\mathbf{Q}$ and $\mathbf{R}$ respectively, $\mathcal{N}$ denotes the normal probability distribution, $k$ is the time step index, $n,m,p\in\mathbb{N}$ are dimensions.
The algorithm can be parted into two major steps, prediction and update.
For the prediction part, first so-called sigma-points have to be calculated.
These points are able to accurately capture the  posterior mean and covariance after propagation through the system \eqref{eq:discr-nonlin-model} up to the 3rd order (Taylor series expansion, please refer to \cite{Wan2000-UKF} for details).
In the second part of the prediction the UKF computes the estimate probability distribution (PD) using the sigma-points as follows:

\begin{equation}
\begin{split}
&\text{PD}\left(\hat{\mathbf{x}}_{k|k-1}\big|\mathbf{y}_{1}...\mathbf{y}_{k-1}\right):=\\
&\mathit{\mathcal{N}}\left(\hat{\mathbf{x}}_{k|k-1}\bigg|\overset{2n}{\underset{i=0}{\sum}}\mathcal{W}_{m}^{(i)}\mathbf{\chi}_{k|k-1}^{(i)},\mathbf{P}_{k|k-1}\right).
\end{split}
\label{eq:UKF-predict-PDF}
\end{equation}

In \eqref{eq:UKF-predict-PDF}, $\mathbf{P}_{k|k-1}$ is the \emph{a priori} estimate covariance, and $\chi_{k|k-1}^{(i)}=$$f\left(\chi_{k-1|k-1}^{(i)},\mathbf{u}_{k-1}\right)$ are the sigma\textendash points with the weights $\mathcal{W}_{c}^{(i)}, \mathcal{W}_{m}^{(i)},i=0,...,2n$.
The predicted mean is computed from the sigma\textendash points by the formula:

\[
\hat{\mathbf{x}}_{k|k-1}=\sum_{i=0}^{2n}\mathcal{W}_{m}^{(i)}\chi_{k|k-1}^{(i)}.
\]

The \emph{a priori} estimate covariance is calculated as follows:

\begin{equation}
\begin{split}
	\mathbf{P}_{k|k-1}=\overset{2n}{\underset{i=0}{\sum}} & \mathcal{W}_{c}^{(i)}\left(\chi_{k|k-1}^{(i)}-\hat{\mathbf{x}}_{k|k-1}\right) \cdot \\
	                                                      & \left(\chi_{k|k-1}^{(i)}-\hat{\mathbf{x}}_{k|k-1}\right)^\top+\mathbf{Q},
\end{split}
\label{eq:a-priori-est-cov}
\end{equation}

where $\mathcal{W}_{c}^{(i)},i=0,...,2n$ are weight factors. 


The update step involves recalculating the sigma\textendash points from $\mathit{\mathcal{N}}\left(\hat{\mathbf{x}}_{k|k-1}|\mathbf{P}_{k|k-1}\right)$.
The mean of the predicted output 
\begin{equation*}
\hat{\mathbf{y}}_k=\sum_{i=0}^{2n}\mathcal{W}_{m}^{(i)}h\left(\chi_{k|k-1}^{(i)}\right)
\end{equation*}
and covariance 
\begin{equation*}
\begin{split}
\mathbf{S}_k=\sum_{i=0}^{2n}\mathcal{W}_{c}^{(i)} \left(h\left(\chi_{k|k-1}^{(i)}\right)-\hat{\mathbf{y}}_k\right) \cdot  \\ 
 \left(h\left(\chi_{k|k-1}^{(i)}\right)-\hat{\mathbf{y}}_k\right)^\top+\mathbf{R},
\end{split}
\end{equation*}
as well as the state and output covariance,
\begin{equation*}
\mathbf{C}_k=\sum_{i=0}^{2n}\mathcal{W}_{c}^{(i)} \left(\chi_{k|k-1}^{(i)}-\hat{\mathbf{x}}_{k|k-1}^{(i)}\right) \cdot \left(h\left(\chi_{k|k-1}^{(i)}\right)-\hat{\mathbf{y}}_k\right)^\top
\end{equation*}
are then used to calculate the Kalman gain $\mathbf{K}_k=\mathbf{C}_k\mathbf{S}_k^{-1}$.
The last step is to update estimate mean with
\begin{equation}
\hat{\mathbf{x}}_{k|k}=\hat{\mathbf{x}}_{k|k-1}+\mathbf{K}_k(\mathbf{y}_k-\hat{\mathbf{y}}_k)
\end{equation} 
and the a posteriori covariance now becomes 
\begin{equation}
\mathbf{P}_{k|k}=\mathbf{P}_{k|k-1}-\mathbf{K}_k\mathbf{S}_k\mathbf{K}_k^\top.
\end{equation}

The choice of the state noise covariance $\mathbf{Q}$ is crucial.
An inappropriate choice may result in divergence issues \citep{Fitzgerald1971-KF-divergence}.
It was suggested to introduce an adaptation matrix $\mathbf{A}_{k}$ so that the state noise covariance becomes effectively $\mathbf{A}_{k}\mathbf{Q}$ (for an extensive description, please refer to \cite{Osinenko2014}).
The adaptation matrix is computed by matching the covariance of the true $\mathbf{y}_{k}$ and the estimated output $\hat{\mathbf{y}}_{k}$ in the sense of
\begin{equation}
\arg \min_{\mathbf{A}_{k}} \left( \mathbf{P}_{k|k-1} -	\mathbf{K}_k \bar{\mathbf{S}}_{k} \mathbf{K}_k^\top \right),
\end{equation} 
where the covariance of $\mathbf{y}_{k}$ is computed from a sample of some size $M$:
\begin{equation}
\bar{\mathbf{S}}_{k}=\frac{1}{M-1} \sum^{k}_{i=k-M+1}(y_i-\hat{y}_i)(y_i-\hat{y}_i)^\top.
\end{equation}
The adaptation of the state noise covariance $\mathbf{Q}$ helps avoid estimate divergence, but it may result in a too noisy estimate. 
This is due to the fact that the noise covariance of the KF defines the tracking strength: the higher $\mathbf{Q}$ is, the more noisy the estimate becomes, but large $\mathbf{Q}$ is needed to avoid divergence. 
The smaller $\mathbf{Q}$ is, the more smooth the estimate is, but there is a risk of divergence. 
To balance these effects, it was suggested to introduce a fuzzy--logic system (FLS) to additionally supervise the UKF \citep{Abdelnour1993-fuzzy-KF}. 
If the vehicle undergoes a phase of intense dynamics, the FLS adaptation factor is set high. 
If the dynamics are steady, the FLS adaptation factor is small. 
The adaption of $\mathbf{Q}$ aids the user by taking care of UKF tuning in an automatic manner.
The resulting filter is called AUKF-FS.
Such a fuzzy classification of dynamics was used in electrified mobile machinery in \cite{Osinenko2015_fuzzy_power_management} and \cite{Osinenko2017}. 


%

According to the vehicle dynamics model of Section~\ref{sec:traction_dynamics}, the state vector for the AUKF-FS consists of the wheel speeds, vehicle ground speed $v$, adhesion coefficients for each wheel and the soil deformation rolling resistance coefficient:

\[
\mathbf{x}=\left(\begin{array}{ccccc}
\omega_\mathrm{{w1}},\dots\omega_\mathrm{{w4}}, & v, & \mu_{1},\dots\mu_{4}, & \rho_\mathrm{{s}}\end{array}\right)^\top.
\]

The wheel speeds and vehicle ground speed form the output vector:

\[
\mathbf{y}=\left(\begin{array}{cc}
\omega_\mathrm{{w1}},\dots,\omega_\mathrm{{w4}}, & v\end{array}\right)^\top.
\]
For the mapping of the identified traction parameters it is assumed that the vehicle has access to GPS, this also allows, in particular, for speed measurement.
Different means are possible for that sake \eg revolution counter on non-drive wheels, a radar etc.

The input vector includes the drive torques, front vertical force $F_\mathrm{zf}$ and longitudinal component of the drawbar pull:

\[
\mathbf{u}=\left(\begin{array}{c}
M_\mathrm{{d1}},\dots M_\mathrm{{d4}},F_\mathrm{{zf}},F_\mathrm{{dx}}\end{array}\right)^\top.
\]

Drive torque can be determined in hydraulic and electrical drive trains, while for mechanical drive trains the measurement/estimation is more elaborate.
Front wheel vertical force can be measured in the suspension and the rear wheel vertical force can then be calculated using vehicle parameters.
Draft force measurement $F_\mathrm{dx}$ can be obtained through magnetoelastic sensors or strain gauges installed in load pins.

The tire deformation rolling resistance coefficients $\rho_\mathrm{t}$ are assumed as fixed parameters. 
The dynamical rolling radii $r_{d1},\dots r_{d4}$ are computed using \eqref{eq:roll_radii} and \eqref{eq:roll_radii_deformation}.
Propagation of the sigma points through the model formed by \eqref{eq:torque_balance} and \eqref{eq:vehicle_dynamics} is performed using the fourth\textendash order Runge\textendash Kutta method.
The dynamics of unknown parameters and rear vertical force $\mu_{1},\dots\mu_{4},\rho_\mathrm{{s}},F_\mathrm{{zr}}$ are unknown.
It is assumed that they  do not change during one integration step so that their dynamics are neglected.
More details are given in \cite{Osinenko2015a}.
It can be easily checked that the overall system is observable.
Two variables, $\omega_{w}$ and $v$, are directly measured.
The other two states can be uniquely calculated at each time step using equations \eqref{eq:torque_balance} and \eqref{eq:vehicle_dynamics}.
The next section focuses specifically on the identification of the $\mu(s)$-characteristic.
%
%
%



\subsection{Adhesion characteristic identification}

With the state observer from the previous section the parameters $\mu$ and $\rho_\mathrm{s}$ are estimated, which provides relevant information of ground conditions. 
As was stated above, the adhesion coefficient is a function of slip (Fig.~\ref{fig:EE_AC_slip_graph}).
The shape of the respective curves changes depending on different soil surfaces.
The Pacejka's empirical models \citep{Pacejka2006-veh-dyn} are usually used for the $\mu(s)$-curve.
Here, it is modified to give
\begin{equation}\label{eq:mu(s)_empirical}
\mu(s) = a - p a \exp^{\alpha_1s} -  a(1-p) \exp^{\alpha_2s},
\end{equation}
where $a,p,\alpha_1,\alpha_2$ are the $\mu$-model parameters (refer to \cite[Fig. 8]{Osinenko2017} for example of parameter values).
For alternative models, refer to \cite{Pacejka2006-veh-dyn,Schreiber2007}. 

\begin{figure}[h]
	\centering
	\includegraphics[width=0.8\linewidth]{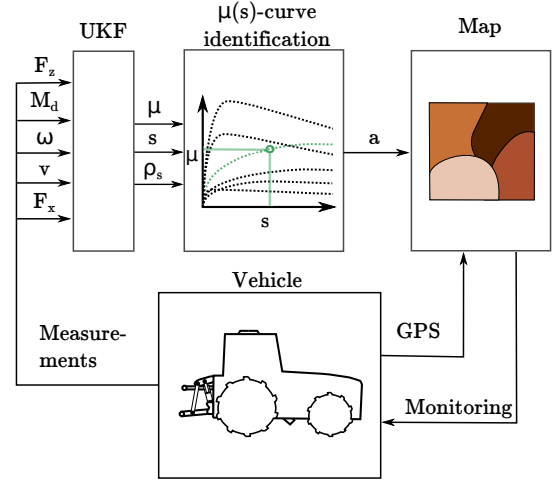}
	\caption{Flowchart of identification and mapping.}
	\label{fig:program_flow}
\end{figure}

\begin{figure}[h]
	\centering
	\includegraphics[width=0.88\linewidth]{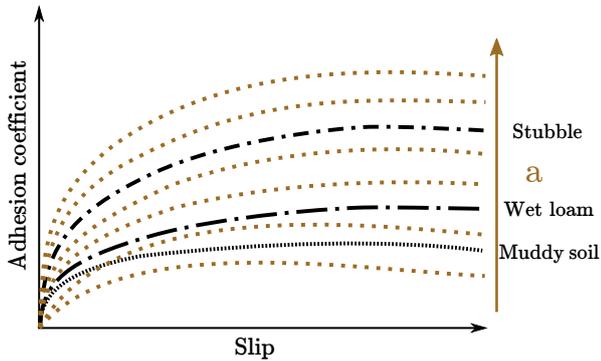}
	\caption{Three similar ground types are displayed. Parameters for \eqref{eq:mu(s)_empirical} are fitted to "stubble". The colored curves are generated by changing one parameter $a$.}
	\label{fig:mu_char}
\end{figure}

Fig.~\ref{fig:program_flow} illustrates this works identification and mapping algorithm schematically.
In the first step of the $\mu(s)$-characteristic identification, a set of typical $\mu$-curves, whose parameters $a,p,\alpha_1,\alpha_2$ are known, is generated. 
It was observed that for similar kinds of surface \eg road or soil, the three parameters  $p,\alpha_1,\alpha_2$ may be fix, while only the parameter $a$ can be varied to cover a spectrum of ground conditions.
This parameter is used in the second step to generate a set of shape-similar $\mu(s)$-curves with only one changing parameter $a$, see Fig.~\ref{fig:mu_char}.

As mentioned above, the AUKF-FS identifies $\mu$, $\rho_\mathrm{s}$ and $s$.
With the estimated pair of $\mu$ and $s$ and the known parameters $p,\alpha_1,\alpha_2$ the parameter $a$ can be calculated.


Now, as the identification measures are done, proceed to the mapping of ground condition parameters.

\section{Mapping} \label{sec:mapping}

The identified parameters are determined at certain discrete instances during the vehicle motion on the working area \eg a field.
Since soil properties are usually similar in a vicinity, interpolation is suggested to generate a map from those discrete estimates.	 
The parameters $p_\mathrm{soil}={a,p,\alpha_1,\alpha_2,\rho_{\mathrm{s}}}$ are known and connected to their respective coordinates, which can be measured through \eg a GPS.
The map is initialized as an empty matrix.
The fetched GPS coordinates are first transformed in such a way that the origin of the measurement coordinate system aligns with the first entry in the map-matrix.
Depending on the resolution of measurements and of the map, there are cases where multiple measurements are available for the same map entry.
In this case, saving the mean value is a reasonable choice.
Since heavy-duty vehicles usually operate at comparably slow speeds, saving measurements at a rate of 10\,Hz is used in this study.
The map-matrix $M$ has the dimensions $w,l,h$, where $w$ is width, $l$ is length and $h$ is the number of parameters in  $p_\mathrm{soil}$
\begin{equation}
	M_{i,j}=p_{\mathrm{soil}\;i,j},
\end{equation}
where $i$ and $j$ denote the matrix indices.

The next step involves the interpolation/extrapolation of the data into the close vicinity of each non-empty map entry.
First, three search thresholds $\varepsilon_\mathrm{low}>\varepsilon_\mathrm{mid}>\varepsilon_\mathrm{high}$ with weights $w_\mathrm{low}<w_\mathrm{mid}<w_\mathrm{high}$ are introduced.
The distance $d$ between two entries in the map-matrix $M_{i,j}$ and $M_{e,f}$ is defined as 
\begin{equation*}
d=|e-i|+|f-j| 
\end{equation*}
which is also known as the Manhatten distance.
Starting with $M_{i,j}=M_{1,1}$, the mean of all entries with a distance $\varepsilon_\mathrm{low}\geq d>\varepsilon_\mathrm{mid}$ is calculated and weighted with $w_\mathrm{low}$.
This is repeated for $w_\mathrm{mid}$ and $w_\mathrm{high}$.
The sum of all weighted mean values is the new extrapolated value for the particular entry $M_{i,j}$.
This procedure is repeated for the whole map and every parameter in $p_\mathrm{soil}$.



\begin{figure}[h]
	\centering
	\includegraphics[width=0.9\linewidth]{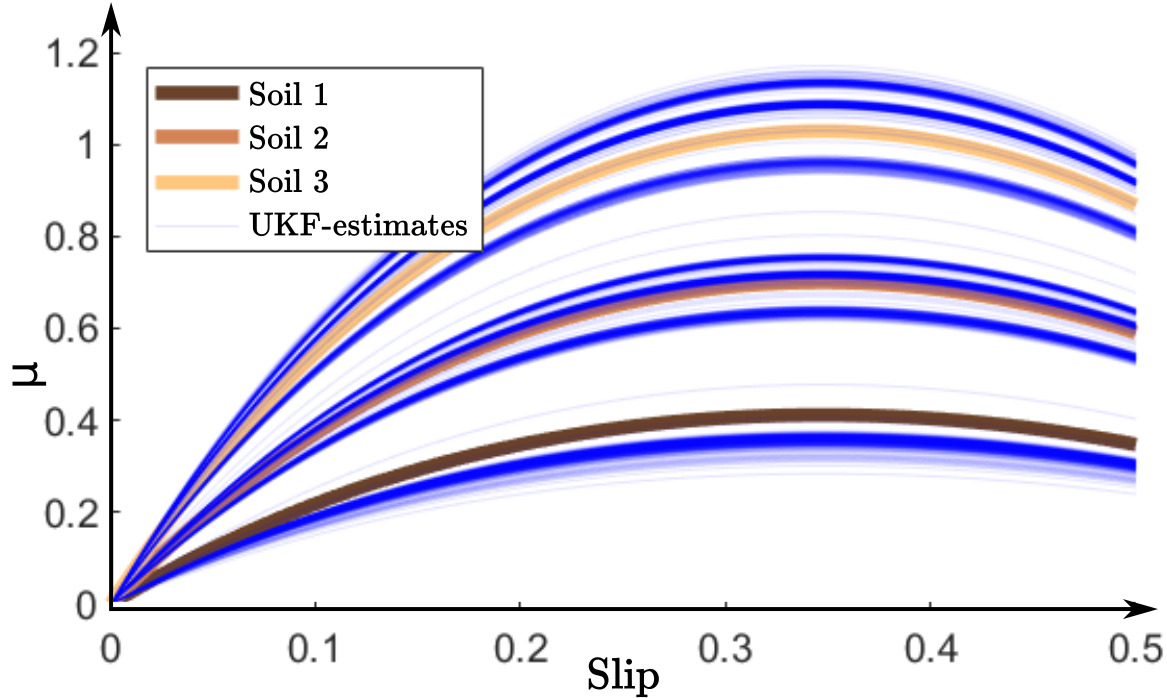}
	\caption{The brown lines represent the true traction characteristics in the simulation. The blue lines show the characteristics identified by the AUKF-FS.}
	\label{fig:mugainmap}
\end{figure}

\section{Results and discussion} \label{sec:results}

The model equations and algorithms introduced in the previous sections were implemented in a Matlab/Simulink environment.
Measurement noise is simulated as white noise.
For the case study, a model of an electrified tractor was used (please refer to \cite{Osinenko2014} 
for comprehensive details). 
It has 80\,kW total drive-train power, 6300\,kg unloaded mass including four wheels with 160\,kg each.
The ground is chosen as a flat plane with three different soil types.
The vehicle accelerates to a desired speed and then cruises at it.
The search thresholds were set as $\varepsilon_\mathrm{low}=10, \; \varepsilon_\mathrm{mid}=5, \; \varepsilon_\mathrm{high}=1.5$ meter and the weights were chosen as $w_\mathrm{low}=0.1, \;w_\mathrm{mid}=0.5, \;w_\mathrm{high}=4$.



\begin{figure}[h]
	\centering
	\includegraphics[width=0.93\linewidth]{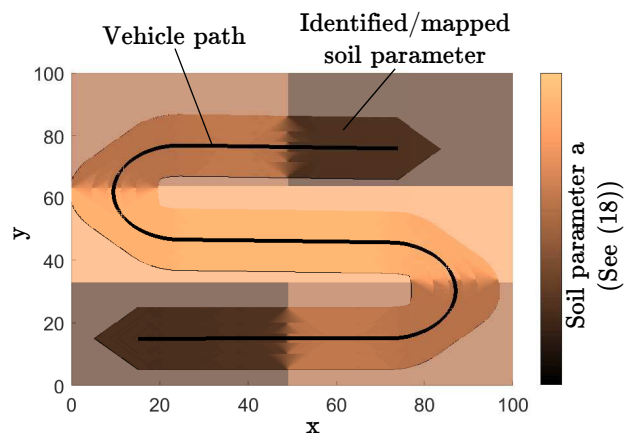}
	\caption{In a simulation a vehicle (black line) drove over a field (transparent background) and simultaneously identified traction parameters (opaque tiles).}
	\label{fig:mapping}
\end{figure}

The adhesion coefficient is calculated using \eqref{eq:mu_physical} and compared to the UKF estimates, see Fig.~\ref{fig:mugainmap}.
A visual evaluation shows a good fit. 
Whenever the soil's traction properties change, the AUKF-FS adapts the estimated values to track the true physical one.
The average absolute estimation error is within 5\%. 
For every ground condition an average representative $\mu(s)$-characteristic was calculated from the UKF-estimates.
These representatives were compared to the true  $\mu(s)$-characteristic using R-squared.
This gave R-squared values of $R^2=0.857$, $R^2=0.996$ and $R^2=0.983$ for soil 1, 2 and 3 respectively.
Fig.~\ref{fig:mapping} shows the map generated from the algorithm suggested in this work.
The background and the identified tiles align well, the AUKF-FS successfully detects changes in soil properties (see the transition between different soils) and identifies parameters reasonably well.

\section{Conclusion} \label{sec:conclusion}

This work demonstrated a method of identification of ground condition parameters combined with their mapping.
These parameters are crucial in determining optimal set-point for operation of heavy-duty vehicles.
A case study with a single-wheel drive moving on a field with three different soil types showed promising capabilities of the mapping algorithm.
The latter may be used \eg in intelligent traction control algorithms or condition monitoring systems.




\begin{ack}
This research is founded by the Saxon Ministery of Science and Art and the 'Sächsische Aufbaubank (SAB)', SAB-project number 100333816.
\end{ack}

\bibliography{bibtex/ctrl-general,bibtex/ctrl-misc,bibtex/electrical-engineering,bibtex/fuzzy-logic,bibtex/identification-misc,bibtex/Kalman-filter,bibtex/measurement,bibtex/motion-dynamics,bibtex/optimization,bibtex/signal-processing,bibtex/slip-ctrl,bibtex/traction-prediction,bibtex/tractors-misc,bibtex/vehicle-identification,bibtex/stochastics}     

\begin{thebibliography}{27}
\providecommand{\natexlab}[1]{#1}
\providecommand{\url}[1]{\texttt{#1}}
\providecommand{\urlprefix}{URL }
\expandafter\ifx\csname urlstyle\endcsname\relax
  \providecommand{\doi}[1]{doi:\discretionary{}{}{}#1}\else
  \providecommand{\doi}{doi:\discretionary{}{}{}\begingroup
  \urlstyle{rm}\Url}\fi

\bibitem[{Abdelnour et~al.(1993)Abdelnour, Chand, and
  Chiu}]{Abdelnour1993-fuzzy-KF}
Abdelnour, G., Chand, S., and Chiu, S. (1993).
\newblock {A}pplying fuzzy logic to the {Kalman} filter divergence problem.
\newblock In \emph{Proceedings of the International Conference on Systems, Man
  and Cybernetics 'Systems Engineering in the Service of Humans'}, 630--635.

\bibitem[{Alexander et~al.(2018)Alexander, Sciancalepore, and
  Vacca}]{Addison2018}
Alexander, A., Sciancalepore, A., and Vacca, A. (2018).
\newblock {Online Controller Setpoint Optimization for Traction Control Systems
  Applied to Construction Machinery}.
\newblock In \emph{Fluid Power Systems Technology}, volume BATH/ASME 2018
  Symposium on Fluid Power and Motion Control.

\bibitem[{Battiato and Diserens(2017)}]{Battiato2017}
Battiato, A. and Diserens, E. (2017).
\newblock Tractor traction performance simulation on differently textured soils
  and validation: A basic study to make traction and energy requirements
  accessible to the practice.
\newblock \emph{Soil and Tillage Research}, 166, 18 -- 32.

\bibitem[{Brixius(1987)}]{Brixius1987}
Brixius, W. (1987).
\newblock {T}raction prediction equations for bias ply tires.
\newblock \emph{ASAE Paper}, 87, 162.

\bibitem[{Fitzgerald(1971)}]{Fitzgerald1971-KF-divergence}
Fitzgerald, R. (1971).
\newblock {D}ivergence of the {Kalman} filter.
\newblock \emph{IEEE Transactions on Automatic Control}, 16(6), 736--747.

\bibitem[{Guskov et~al.(1988)Guskov, Velev, Atamanov, Bocharov, Ksenevich, and
  Solonsky}]{Guskov1988}
Guskov, V.V., Velev, N.N., Atamanov, Y.E., Bocharov, N.F., Ksenevich, I.P., and
  Solonsky, A.S. (1988).
\newblock \emph{{T}raktory: {T}eoriya: {U}chebnik dlya {S}tudentov {V}uzov, po
  {S}pecialnosti "{A}vtomobili i {T}raktory". [{T}ractors. {T}heory. {T}extbook
  for {S}tudents of {H}igher {E}ducational {I}nsitutions {M}ajoring in
  {A}utomotive and {T}ractor {T}echnology (in {R}ussian)]}.
\newblock Moscow: Mashinostroenie.

\bibitem[{Hamann et~al.(2014)Hamann, Hedrick, Rhode, and
  Gauterin}]{Hamann2014-adhesion-est-UKF}
Hamann, H.F., Hedrick, J.K., Rhode, S., and Gauterin, F. (2014).
\newblock Tire force estimation for a passenger vehicle with the unscented
  kalman filter.
\newblock In \emph{Intelligent Vehicles Symposium Proceedings, 2014 IEEE},
  814--819. IEEE.

\bibitem[{{I}shikawa et~al.(2012){I}shikawa, {N}ishi, {O}kabe, and
  {Y}agi}]{Ishikawa2012}
{I}shikawa, S., {N}ishi, E., {O}kabe, N., and {Y}agi, K. (2012).
\newblock {D}ata processing: vehicles, navigation, and relative location
  vehicle control, guidance, operation, or indication construction or
  agricultural-type vehicle (e.g., crane, forklift). {J}apanese patent
  {AA}01{B}7100{FI}.

\bibitem[{Kim and Lee(2018)}]{Kim2018}
Kim, J. and Lee, J. (2018).
\newblock Traction-energy balancing adaptive control with slip optimization for
  wheeled robots on rough terrain.
\newblock \emph{Cognitive Systems Research}, 49, 142 -- 156.

\bibitem[{{Novak} and {Va\v{s}ak}(2018)}]{Novak2018}
{Novak}, H. and {Va\v{s}ak}, M. (2018).
\newblock Energy-efficient train traction control on complex rail
  configurations.
\newblock In \emph{2018 26th Mediterranean Conference on Control and Automation
  (MED)}, 1--9.

\bibitem[{Osinenko et~al.(2015{\natexlab{a}})Osinenko, Geißler, and
  Herlitzius}]{Osinenko2015_fuzzy_power_management}
Osinenko, P., Geißler, M., and Herlitzius, T. (2015{\natexlab{a}}).
\newblock Fuzzy-logic assisted power management for electrified mobile
  machinery.
\newblock \emph{Neurocomputing}, 170, 439 -- 447.

\bibitem[{Osinenko et~al.(2016)Osinenko, Geissler, Herlitzius, and
  Streif}]{Osinenko2016}
Osinenko, P., Geissler, M., Herlitzius, T., and Streif, S. (2016).
\newblock Experimental results of slip control with a fuzzy-logic-assisted
  unscented kalman filter for state estimation.
\newblock In \emph{2016 IEEE International Conference on Fuzzy Systems,
  FUZZ-IEEE 2016}, 501--507.

\bibitem[{Osinenko et~al.(2014)Osinenko, Geissler, and
  Herlitzius}]{Osinenko2014}
Osinenko, P., Geissler, M., and Herlitzius, T. (2014).
\newblock Adaptive unscented kalman filter with a fuzzy supervisor for
  electrified drive train tractors.
\newblock In \emph{IEEE International Conference on Fuzzy Systems}.

\bibitem[{Osinenko and Streif(2017)}]{Osinenko2017}
Osinenko, P. and Streif, S. (2017).
\newblock {O}ptimal traction control for heavy-duty vehicles.
\newblock \emph{{C}ontrol Engineering Practice}, 69, 99 -- 111.

\bibitem[{Osinenko et~al.(2015{\natexlab{b}})Osinenko, Geissler, and
  Herlitzius}]{Osinenko2015a}
Osinenko, P.V., Geissler, M., and Herlitzius, T. (2015{\natexlab{b}}).
\newblock A method of optimal traction control for farm tractors with feedback
  of drive torque.
\newblock \emph{Biosystems Engineering}, 129, 20 -- 33.

\bibitem[{{P}acejka(2006)}]{Pacejka2006-veh-dyn}
{P}acejka, H.B. (2006).
\newblock \emph{{T}yre and {V}ehicle {D}ynamics}.
\newblock Automotive engineering. Butterworth-Heinemann.

\bibitem[{Pentos and Pieczarka(2017)}]{Pentos2017}
Pentos, K. and Pieczarka, K. (2017).
\newblock Applying an artificial neural network approach to the analysis of
  tractive properties in changing soil conditions.
\newblock \emph{Soil and Tillage Research}, 165, 113 -- 120.

\bibitem[{{R}ajamani et~al.(2012){R}ajamani, {P}hanomchoeng, {P}iyabongkarn,
  and {L}ew}]{Rajamani2012-adhesion-est}
{R}ajamani, R., {P}hanomchoeng, G., {P}iyabongkarn, D., and {L}ew, J.Y. (2012).
\newblock {A}lgorithms for {R}eal-{T}ime {E}stimation of {I}ndividual {W}heel
  {T}ire-{R}oad {F}riction {C}oefficients.
\newblock \emph{IEEE/ASME Trans. Mechatron.}, 17(6), 1183--1195.

\bibitem[{{Reichensd{\"o}rfer} et~al.(2018){Reichensd{\"o}rfer}, {Odenthal},
  and {Wollherr}}]{Reichensdoerfer2018}
{Reichensd{\"o}rfer}, E., {Odenthal}, D., and {Wollherr}, D. (2018).
\newblock On the stability of nonlinear wheel-slip zero dynamics in traction
  control systems.
\newblock \emph{IEEE Transactions on Control Systems Technology}, 1--16.

\bibitem[{{S}chreiber and {K}utzbach(2007)}]{Schreiber2007}
{S}chreiber, M. and {K}utzbach, H. (2007).
\newblock {C}omparison of different zero-slip definitions and a proposal to
  standardize tire traction performance.
\newblock \emph{Journal of Terramechanics}, 44(1), 75--79.

\bibitem[{Schreiber et~al.(2008)Schreiber, Kutzbach et~al.}]{Schreiber2008}
Schreiber, M., Kutzbach, H., et~al. (2008).
\newblock {I}nfluence of soil and tire parameters on traction.
\newblock \emph{Research in Agricultural Engineering}, 54, 43--49.

\bibitem[{S{\"o}hne(1964)}]{Sohne1964}
S{\"o}hne, W. (1964).
\newblock {A}llrad- oder {H}interradantrieb bei {A}ckerschleppern hoher
  {L}eistung [{A}ll whell or rear wheel drive train of farm tractors with high
  eingine power (in {G}erman)].
\newblock \emph{Grundlagen der Landtechnik [Basics of agricultural
  engineering]}, 20, 44--52.

\bibitem[{Turnip and Fakhrurroja(2013)}]{Turnip2013-adhesion-est-EKF}
Turnip, A. and Fakhrurroja, H. (2013).
\newblock Estimation of the wheel-ground contacttire forces using extended
  kalman filter.
\newblock \emph{International Journal of Instrumentation Science}, 2(2),
  34--40.

\bibitem[{Van Der~Merwe et~al.(2004)Van Der~Merwe, Wan, and
  Julier}]{VanDerMerwe2004-UKF}
Van Der~Merwe, R., Wan, E.A., and Julier, S. (2004).
\newblock {S}igma-point {K}alman filters for nonlinear estimation and
  sensor-fusion. {A}pplications to integrated navigation.
\newblock In \emph{Proceedings of the AIAA Guidance, Navigation \& Control
  Conference}, 16--19.

\bibitem[{Wan and Van Der~Merwe(2000)}]{Wan2000-UKF}
Wan, E.A. and Van Der~Merwe, R. (2000).
\newblock {T}he unscented {K}alman filter for nonlinear estimation.
\newblock In \emph{Proceedings of the IEEE Symposium on Adaptive Systems for
  Signal Processing, Communications, and Control 2000. AS-SPCC}, 153--158.

\bibitem[{Wang et~al.(2004)Wang, Alexander, and
  Rajamani}]{Wang_FrictionEstimation}
Wang, J., Alexander, L., and Rajamani, R. (2004).
\newblock {Friction Estimation on Highway Vehicles Using Longitudinal
  Measurements }.
\newblock \emph{Journal of Dynamic Systems, Measurement, and Control}, 126(2),
  265--275.

\bibitem[{{Zhe Jiang} et~al.(2007){Zhe Jiang}, {Qi Song}, {Yuqing He}, and
  {Jianda Han}}]{Jiang2007}
{Zhe Jiang}, {Qi Song}, {Yuqing He}, and {Jianda Han} (2007).
\newblock A novel adaptive unscented kalman filter for nonlinear estimation.
\newblock In \emph{2007 46th IEEE Conference on Decision and Control},
  4293--4298.

\end{thebibliography}







\end{document}